\documentclass[11pt]{article}

\usepackage[final]{acl}

\usepackage{times}
\usepackage{latexsym}

\usepackage[T1]{fontenc}

\usepackage[utf8]{inputenc}

\usepackage{microtype}

\usepackage{inconsolata}

\usepackage{graphicx}
\usepackage{amsmath}

\usepackage{booktabs}
\usepackage{tcolorbox}
\usepackage{amsfonts}
\usepackage{multirow}

\title{
  \raisebox{-0.31\height}{\includegraphics[width=1cm, height=1cm]{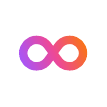}}
  FlashLabs Chroma 1.0: A Real-Time End-to-End Spoken Dialogue Model with Personalized Voice Cloning
}

\author{
 \textbf{Tanyu Chen\textsuperscript{*}},
 \textbf{Tairan Chen\textsuperscript{*}},
 \textbf{Kai Shen\textsuperscript{*}},
 \textbf{Zhenghua Bao\textsuperscript{*,$\dagger$}},
\\
 \textbf{Zhihui Zhang},
 \textbf{Man Yuan},
 \textbf{Yi Shi\textsuperscript{$\dagger$}}
\\
    FlashLabs
\\
 {
 \small
  \textbf{Correspondence:} 
  \href{mailto:zhenghua.bao@flashlabs.ai}{zhenghua.bao@flashlabs.ai},
  \href{mailto:chroma@flashlabs.ai}{chroma@flashlabs.ai}
 }
}

\begin{document}
\maketitle

\renewcommand{\thefootnote}{*}
\footnotetext{Equal contribution.}
\renewcommand{\thefootnote}{$\dagger$}
\footnotetext{Corresponding author.}
\renewcommand{\thefootnote}{\arabic{footnote}} 

\begin{abstract}
Recent end-to-end spoken dialogue systems leverage speech tokenizers and neural audio codecs to enable LLMs to operate directly on discrete speech representations. However, these models often exhibit limited speaker identity preservation, hindering personalized voice interaction. In this work, we present \textbf{Chroma 1.0}, the first open-source, real-time, end-to-end spoken dialogue model that achieves both low-latency interaction and high-fidelity personalized voice cloning. Chroma achieves sub-second end-to-end latency through an interleaved text-audio token schedule (1:2) that supports streaming generation, while maintaining high-quality personalized voice synthesis across multi-turn conversations. Our experimental results demonstrate that Chroma achieves a \textbf{10.96\%} relative improvement in speaker similarity over the human baseline, with a Real-Time Factor of \textbf{0.43}, while maintaining strong reasoning and dialogue capabilities. Our code and models are publicly available at \href{https://github.com/FlashLabs-AI-Corp/FlashLabs-Chroma}{GitHub} and \href{https://huggingface.co/FlashLabs/Chroma-4B}{HuggingFace}.
\end{abstract}

\begin{figure*}[!t]
    \centering
    \includegraphics[width=0.88\textwidth]{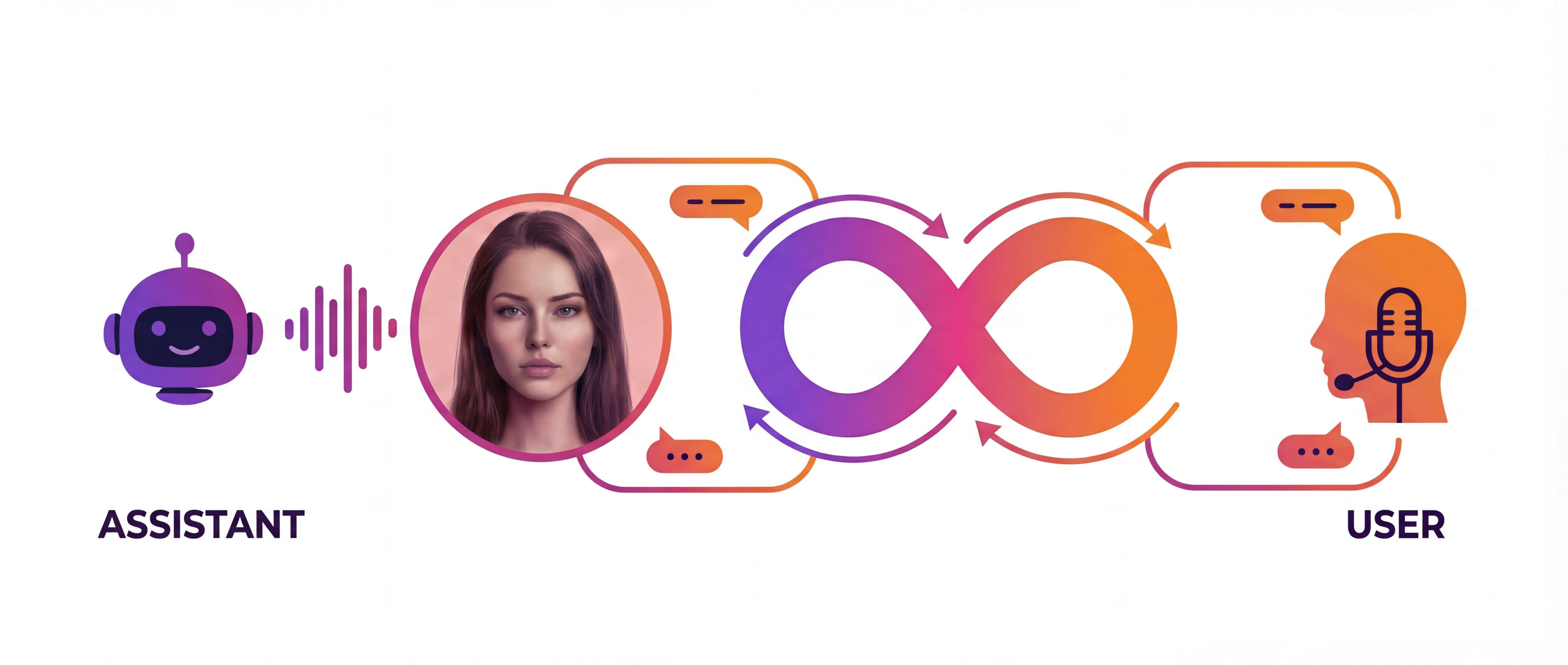}
    \caption{
        System workflow of \textbf{Chroma 1.0}. Chroma takes speech as input and produces speech as output, maintaining consistent speaker identity throughout the conversation.
        }
    \label{fig:model_architecture}
\end{figure*}

\section{Introduction}
Building real-time, interactive speech dialogue systems remains challenging. 
Most deployed systems follow a cascaded pipeline: input speech is first transcribed into text via automatic speech recognition (ASR), the transcript is then processed by a downstream large language model (LLM), and the generated response is finally converted back to speech by a text-to-speech (TTS) module. 
While effective for offline or latency-tolerant scenarios, such pipelines incur high end-to-end latency, suffer from error propagation, and tend to lose paralinguistic information such as speaker identity, speaking rate, timbre, emotion, and prosody. 
These limitations become especially limiting in real-time conversational settings, 
where naturalness, responsiveness, and speaker fidelity are essential.

Recent advances in speech tokenization and neural codecs~\cite{hsu2021hubert, défossez2022highfidelityneuralaudio, ji2025wavtokenizerefficientacousticdiscrete} enable LLMs to operate directly on discrete speech representations, giving rise to \emph{speech-to-speech} (S2S) systems that bypass explicit intermediate transcription. 
End-to-end large audio language models (LALMs) emerged following the success of GPT-4o~\cite{openai2024gpt4ocard}, which demonstrated the feasibility of end-to-end speech processing without explicit textual intermediates. Subsequent works~\cite{défossez2024moshispeechtextfoundationmodel, zeng2024glm4voiceintelligenthumanlikeendtoend, xu2025qwen25omnitechnicalreport, xu2025qwen3omnitechnicalreport, nguyen2024spiritlminterleavedspoken, huang2025stepaudioaqaafullyendtoendexpressive,huang2025stepaudiounifiedunderstandinggeneration,  wu2025stepaudio2technicalreport} have adopted this principle, processing and generating speech directly through diverse architectural strategies: interleaved speech-text token sequences~\cite{nguyen2024spiritlminterleavedspoken}, unified audio-text representations~\cite{huang2025stepaudioaqaafullyendtoendexpressive,huang2025stepaudiounifiedunderstandinggeneration,  wu2025stepaudio2technicalreport, zeng2024glm4voiceintelligenthumanlikeendtoend}, or parallel dual-stream architectures that decouple semantic reasoning from acoustic generation~\cite{xu2025qwen25omnitechnicalreport, xu2025qwen3omnitechnicalreport}.

However, existing end-to-end LALMs still face several critical limitations in both speech understanding and speech generation.
Early models such as Spirit LM~\cite{nguyen2024spiritlminterleavedspoken} and GLM-4-Voice~\cite{zeng2024glm4voiceintelligenthumanlikeendtoend} primarily focus on aligning semantic content from speech to text, often showing limited ability to capture paralinguistic cues that are essential for natural interaction.
While more recent audio-understanding models like Qwen2-Audio~\cite{chu2024qwen2audiotechnicalreport} and Kimi-Audio~\cite{kimiteam2025kimiaudiotechnicalreport} can comprehend such paralinguistic information, they generate only textual outputs, failing to leverage this understanding for expressive speech generation.
Moreover, current S2S systems typically prioritize dialogue quality over personalized voice fidelity. While speech systems capable of voice cloning, such as 
VALL-E~\cite{wang2023neuralcodeclanguagemodels}, the CosyVoice 
series~\cite{du2024cosyvoicescalablemultilingualzeroshot, 
du2024cosyvoice2scalablestreaming, du2025cosyvoice3inthewildspeech}, 
and industrial solutions (e.g., Qwen-TTS\footnote{\url{https://qwenlm.github.io/blog/qwen-tts/}} 
and ElevenLabs\footnote{\url{https://elevenlabs.io/voice-cloning}}), achieve high-quality speaker adaptation, they lack real-time streaming 
capabilities with consistent voice cloning across multi-turn conversations. Conversely, real-time dialogue models~\cite{défossez2024moshispeechtextfoundationmodel, xie2024miniomnilanguagemodelshear, xie2024miniomni2opensourcegpt4ovision, fang2025llamaomniseamlessspeechinteraction} sacrifice fine-grained speaker control for low latency.

To address these limitations, we propose \textbf{Chroma 1.0}, the first open-source, real-time end-to-end spoken dialogue model that achieves both low-latency interaction and high-fidelity personalized voice cloning. Our main contributions are as follows:
\begin{itemize}
    \item A streaming architecture that tightly couples speech understanding with speech generation through semantic state representations, enabling sub-second end-to-end latency.
    \item High-fidelity voice cloning that conditions the generation model on audio embeddings from just a few seconds of reference audio, achieving 10.96\% relative improvement in speaker similarity over human baseline.
    \item An interleaved text-audio token schedule (1:2) that enables synchronized generation of acoustic codes with incremental text output, achieving real-time streaming synthesis.
    \item Strong reasoning and dialogue capabilities with only 4B parameters, demonstrating competitive performance across understanding, reasoning, and oral conversation tasks.
\end{itemize}

We release the complete codebase, training pipeline, and pretrained model weights to support reproducibility and further research\footnote{\url{https://github.com/FlashLabs-AI-Corp/FlashLabs-Chroma}} \footnote{\url{https://huggingface.co/FlashLabs/Chroma-4B}}.

\begin{figure*}[!t]
    \centering
    \includegraphics[width=\textwidth]{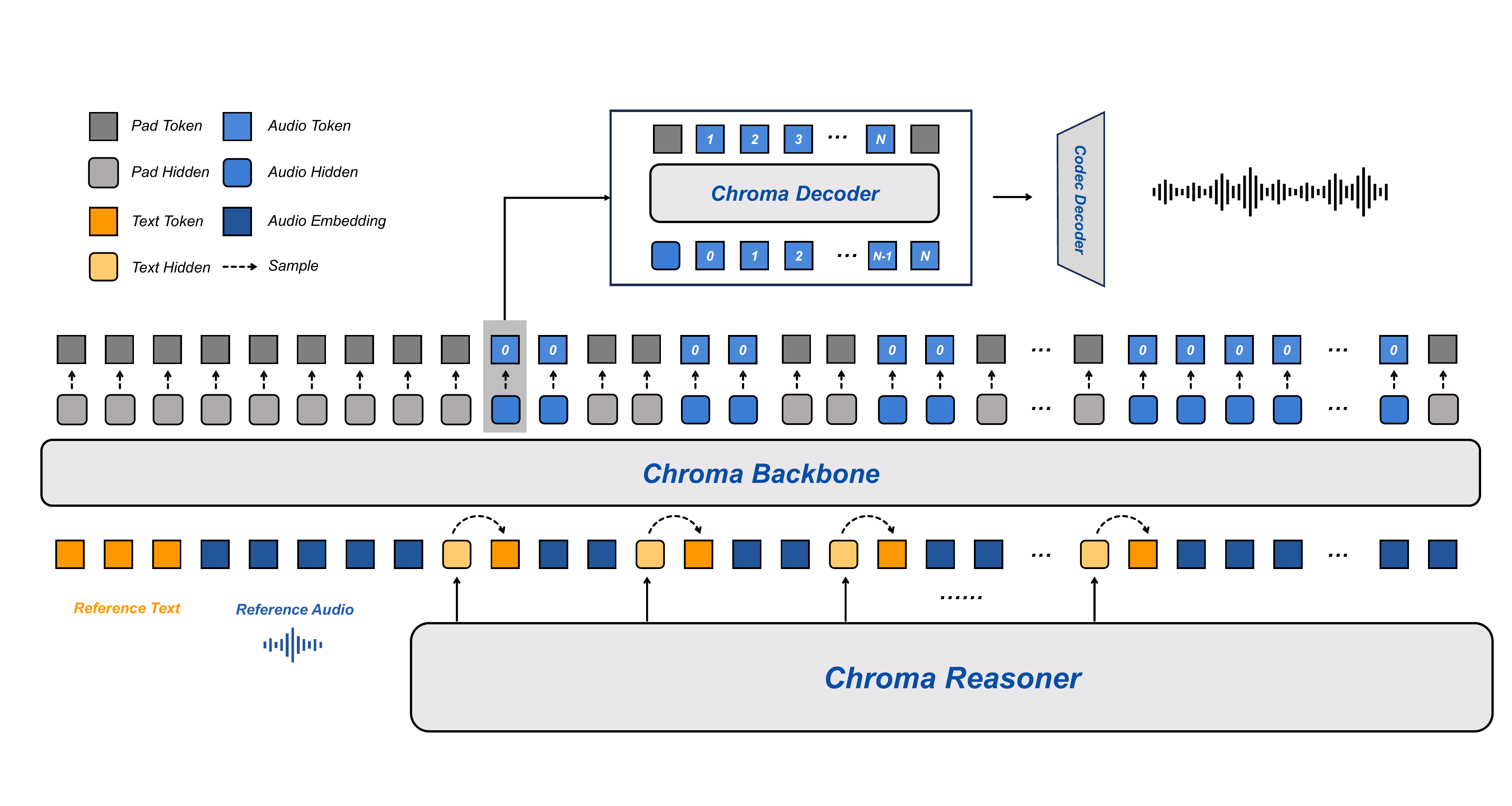}
    \caption{
        Overall architecture of \textbf{Chroma 1.0}. The \textbf{Reasoner} outputs text
        tokens and hidden states. 
        These form an interleaved text–audio embedding sequence (1:2) consumed by the
        \textbf{Backbone} to generate coarse acoustic codes $c_t^{0}$ and hidden states
        $h_t$. The \textbf{Decoder} predicts the remaining RVQ levels
        $c_t^{1:N-1}$, and the \textbf{Codec Decoder} reconstructs the full codebook sequence
        into a continuous waveform, enabling high-fidelity S2S generation.
        }
    \label{fig:model_architecture}
\end{figure*}

\section{Related Works}
\subsection{Cascaded Speech--Text Pipelines vs.\ End-to-End S2S Conversation}
Cascaded approaches remain widely deployed due to the maturity of ASR, LLM, and TTS modules, as well as established engineering practices~\cite{zhang2023speechgpt, radford2023robust, wang2023neuralcodeclanguagemodels,chu2023qwen, tan2024naturalspeech, huang2024audiogpt, lin2024advancing, fang2025llamaomniseamlessspeechinteraction}. They offer flexibility but accumulate latency and errors, and often discard paralinguistic cues once speech is reduced to text, limiting downstream speaker fidelity and emotional expressiveness.
In contrast, end-to-end S2S dialogue systems avoid explicit transcription by learning to map directly between discrete speech units.
SeamlessM4T~\cite{communication2023seamlessm4tmassivelymultilingual} established a multilingual S2S baseline, while GPT-4o~\cite{openai2024gpt4ocard} demonstrated the commercial viability of native audio processing without intermediate text conversions.
Recent research efforts have explored diverse architectural strategies for end-to-end speech modeling.
Moshi~\cite{défossez2024moshispeechtextfoundationmodel} introduces a full-duplex streaming architecture for real-time dialogue, while Spirit LM~\cite{nguyen2024spiritlminterleavedspoken} employs interleaved speech-text token sequences to maintain multimodal alignment.
GLM-4-Voice~\cite{zeng2024glm4voiceintelligenthumanlikeendtoend} and the early Step-Audio series~\cite{huang2025stepaudioaqaafullyendtoendexpressive,huang2025stepaudiounifiedunderstandinggeneration} adopt unified token representations that map speech and text into a shared vocabulary space, with Step-Audio-2~\cite{wu2025stepaudio2technicalreport} further advancing generation quality using retrieval-augmented generation (RAG) and reinforcement learning techniques.
Qwen2.5-Omni~\cite{xu2025qwen25omnitechnicalreport} and Qwen3-Omni~\cite{xu2025qwen3omnitechnicalreport} propose parallel dual-stream (Thinker-Talker) architectures that decouple semantic reasoning from acoustic generation. Specifically, Qwen3-Omni employs a multi-codebook token prediction (MTP) module in the Talker to predict residual codebooks per frame.
Models focused primarily on audio understanding—such as Qwen2-Audio~\cite{chu2024qwen2audiotechnicalreport} and Kimi-Audio~\cite{kimiteam2025kimiaudiotechnicalreport}—generate textual responses, demonstrating the capability to comprehend paralinguistic information from speech inputs.

\subsection{Voice Cloning and Personalized Speech Synthesis}

A major advancement in voice cloning came with \emph{neural codec language models} (NCLMs), which cast TTS as discrete acoustic token generation. VALL-E \citep{wang2023neuralcodeclanguagemodels} demonstrated that modeling EnCodec codes with a large conditional language model yields natural, expressive zero-shot TTS from approximately 3 seconds of reference audio while preserving paralinguistic cues. VALL-E X \citep{zhang2023speakforeignlanguagesvoice} extends this capability across languages. Diffusion and flow-matching approaches further advance fidelity and control. StyleTTS-2 \citep{li2023styletts2humanleveltexttospeech} introduces style diffusion achieving strong naturalness and zero-shot adaptation, NaturalSpeech 3 \citep{ju2024naturalspeech3zeroshotspeech} factorizes content, prosody, and timbre for state-of-the-art similarity, and Voicebox \citep{le2023voiceboxtextguidedmultilingualuniversal} demonstrates high-speed non-autoregressive flow-matching. Open-source systems like CosyVoice series \citep{du2024cosyvoicescalablemultilingualzeroshot, du2024cosyvoice2scalablestreaming, du2025cosyvoice3inthewildspeech}, and OpenVoice \citep{qin2024openvoiceversatileinstantvoice} demonstrate practical cloning with improved content consistency and fine-grained control, with later iterations adding streaming support. Commercial platforms (e.g. Elevenlabs) further validate few-second cloning at production scale.

\section{Model Architecture}

Chroma 1.0 is built upon an end-to-end spoken dialogue architecture designed to achieve high-quality, real-time voice interaction through tight coupling of speech perception and language understanding. As illustrated in Fig.~\ref{fig:model_architecture}, the system consists of two integrated subsystems: (i) the \textbf{Chroma Reasoner}, responsible for multimodal input comprehension and textual response generation; and (ii) a speech synthesis pipeline composed of the \textbf{Chroma Backbone} for acoustic modeling, the \textbf{Chroma Decoder} for audio codebook decoding, and the \textbf{Chroma Codec Decoder} for waveform reconstruction.

\subsection{Chroma Reasoner}

The Chroma Reasoner, built upon the Thinker module~\cite{xu2025qwen25omnitechnicalreport}, performs multimodal understanding and semantic representation construction. It processes both text and audio inputs through the standard Qwen2-Audio encoding pipeline~\cite{chu2024qwen2audiotechnicalreport}, generating high-level semantic representations that capture both linguistic content and acoustic characteristics.

The Reasoner employs a cross-modal attention mechanism to fuse text and audio features. The encoded representations are unified into a sequence of hidden states with temporal alignment through Time-aligned Multimodal Rotary Position Embedding (TM-RoPE)~\cite{xu2025qwen25omnitechnicalreport}. This fusion enables the model to leverage prosodic and rhythmic cues from speech alongside textual semantics, enhancing both dialogue understanding and contextual modeling for subsequent speech synthesis.

\subsection{Chroma Backbone}

The Chroma Backbone adopts a 1B-parameter variant of the LLaMA~\cite{touvron2023llama} architecture, designed to generate speech that matches the timbre of a given reference audio. To enable high-fidelity voice cloning, we encode the reference audio and its corresponding transcript into embedding prompts using CSM-1B\footnote{\url{https://github.com/SesameAILabs/csm}} and prepend them to the input sequence, explicitly conditioning the model on the target speaker's acoustic characteristics. During inference, to ensure strict alignment between the audio output and the Reasoner's text modality while maintaining low parameter count, we apply a shared token embedding strategy: the Reasoner's token embeddings and hidden states are fed into the Backbone as unified textual context.

To support efficient streaming generation, we interleave text tokens with audio code tokens $c^{0}$ at a fixed ratio of $1:2$, meaning each text token is paired with two audio codes. This mechanism enables the Backbone to autoregressively generate audio sequences in parallel with the Reasoner's incremental text generation, allowing the system to produce outputs without waiting for complete text sequences. Consequently, the Time-to-First-Token (TTFT) is significantly reduced, enabling improved real-time interaction.

\subsection{Chroma Decoder}

To substantially accelerate inference while preserving generation quality, we introduce a lightweight module—the \textbf{Chroma Decoder}—responsible for generating the remaining acoustic codes ($c_t^{1}, \ldots, c_t^{N-1}$), rather than having the Backbone produce all codebooks directly. The Chroma Decoder is implemented as a variant of the LLaMA architecture with approximately 100M parameters. Unlike the Backbone, this module does not rely on the full history of text inputs or reference audio context; instead, it performs frame-synchronous inference conditioned only on the Backbone outputs at the current time step, thereby greatly reducing computational overhead associated with long-context processing.

Specifically, at each time step $t$, the Chroma Decoder takes as input the hidden-state features $h_t$ and the initial audio codebook $c_t^{0}$ produced by the Backbone, and autoregressively generates the remaining RVQ codebooks $c_t^{i}$ ($i=1,\ldots,N-1$) within each frame using level-specific projection heads conditioned on previously generated levels. This decoupled design not only reduces inference latency but also enables the Chroma Decoder to enrich fine-grained acoustic attributes, such as prosody and articulation details, building upon the coarse semantic representation from the Backbone.

\subsection{Chroma Codec Decoder}

As the final acoustic reconstruction module, the \textbf{Chroma Codec Decoder} maps the discrete codebook sequence into a continuous, high-fidelity speech waveform. At each time step, the module concatenates the coarse codebook ($c^0$) and the refined acoustic codebooks ($c^{1}, \ldots, c^{N-1}$) generated by the Chroma Decoder to form the complete discrete acoustic representation. Architecturally, this module follows the decoder design of the Mimi vocoder~\cite{défossez2024moshispeechtextfoundationmodel} and employs a causal convolutional neural network (Causal CNN), ensuring strict temporal causality during waveform reconstruction to support streaming generation.

To meet real-time interaction requirements, we employ 8 codebooks ($N = 8$). This configuration significantly reduces the autoregressive refinement steps required by the Chroma Decoder, thereby improving inference efficiency.

\subsection{Training Datasets}

Publicly available datasets lack high-quality speech dialogue data that meet our model's requirements for semantic understanding and reasoning capabilities. To address this limitation, we propose a data generation pipeline that leverages the synergistic collaboration between LLMs and TTS systems.

\textbf{Pipeline Workflow.} The generation process consists of two stages: (1) \textit{Text Generation}: User questions are fed into a Reasoner-like LLM module to generate corresponding textual responses. (2) \textit{Speech Synthesis}: The textual responses are synthesized into speech using a TTS system, with timbre characteristics matching the reference audio. This synthesized speech serves as the training target, enabling the Backbone and Decoder modules to learn voice cloning and acoustic modeling.

\subsection{Training Objective}
Our training strategy optimizes two primary components: the Backbone and the Decoder, while keeping the Reasoner frozen as a feature extractor. For each audio-text pair, the Reasoner provides fixed text embeddings and multimodal hidden states that serve as semantic and prosodic conditioning.
The Backbone is trained to autoregressively predict the first layer of coarse acoustic codes ($c^0$). To ensure causal alignment, it attends only to the prefix of the acoustic codes and the corresponding Reasoner representations. This objective enables the model to capture long-term temporal structure and align acoustic generation with the text progression.
The Decoder refines the coarse acoustic representation by predicting the remaining Residual Vector Quantization (RVQ) levels ($c^{1:N-1}$). Conditioned on the Backbone's coarse code and hidden state, the Decoder operates via an intra-frame autoregressive process. This factorization allows it to progressively enhance acoustic fidelity while maintaining consistency with the coarse trajectory established by the Backbone. More details can be found in Appendix~\ref{app:training_obj}.

\section{Experiments}
\subsection{Experimental Setup}

\paragraph{Datasets.}
We evaluate our model on multiple benchmark datasets to assess different aspects of speech generation quality. For general performance evaluation, we use the \textbf{CommonVoice dataset}~\cite{ardila2020common}, which provides diverse speakers and recording conditions for measuring speaker similarity. To assess voice cloning capability, we conduct subjective experiments measuring naturalness and voice fidelity between generated and reference audio. To evaluate the model's ability to understand and reason about input speech, we utilize a subset of \textbf{URO-Bench}~\cite{yan2025uro}, which provides speech-based question-answering tasks requiring semantic comprehension. Unless otherwise specified, all experiments were conducted on an NVIDIA H200 GPU.

\paragraph{Training Configuration.} 
Our implementation is based on PyTorch 2.7.1. We employ the AdamW optimizer~\cite{loshchilov2017decoupled} with a learning rate of $5 \times 10^{-5}$ and a per-device batch size of 4. The model is trained for 100K steps on 8 NVIDIA H200 GPUs (141GB memory each), achieving convergence in approximately 6 hours. Gradient clipping with a maximum norm of 1.0 is applied to ensure training stability.

\paragraph{Evaluation Metrics.} 
We conduct comprehensive evaluation using both objective and subjective metrics, covering speech quality, naturalness, speaker similarity, and system efficiency. For objective evaluation, we employ Speaker Similarity (SIM), Time-to-First-Token (TTFT), and Real-Time Factor (RTF). For subjective evaluation, we conduct Comparative Mean Opinion Score (CMOS) tests through human listeners. Our SIM evaluation is based on the SEED-TTS-EVAL framework\footnote{\url{https://github.com/BytedanceSpeech/seed-tts-eval}}.

\textbf{Speaker Similarity (SIM)} evaluates voice cloning quality by measuring how closely the generated speech matches the target speaker's voice characteristics. We use WavLM-Large~\cite{chen2022wavlm}, a self-supervised speech representation model fine-tuned on speaker verification tasks~\cite{chen2022large}, to extract 192-dimensional speaker embeddings from both generated and reference audio. Similarity is computed as the cosine similarity between these embeddings.

\textbf{Comparative Mean Opinion Score (CMOS)} measures naturalness (NCMOS) and speaker similarity (SCMOS) through pairwise comparisons. For NCMOS, participants compare two systems and select from four options: ``A sounds more natural'', ``B sounds more natural'', ``About the same'', or ``Hard to tell''. For SCMOS, participants first listen to reference audio, then compare which of two synthesized samples better matches the reference speaker's characteristics using the same four-option scale. The order of samples A and B is randomized to avoid position bias. CMOS scores are computed as the mean preference difference, where positive values indicate preference for our system. 

\textbf{Latency Metrics.} We measure system efficiency using two complementary metrics: Time-to-First-Token (TTFT) and Real-Time Factor (RTF). TTFT measures system responsiveness by recording the time elapsed from receiving the input to generating the first audio token, which is critical for natural conversation flow. RTF measures computational efficiency as the ratio of generation time to audio duration. An RTF below 1.0 indicates the system can generate speech faster than real-time playback. Both metrics are essential for evaluating real-time interactive performance, with lower values indicating better efficiency.

\begin{table}[t]
\centering
\begin{tabular}{lc}
\toprule
\textbf{Model} & \textbf{SIM} $\uparrow$ \\
\midrule
Human Baseline  & 0.73 \\
\midrule
F5-TTS & 0.64 \\
Seed-TTS & 0.76 \\
FireRedTTS-2 & 0.66 \\
Step-Audio-TTS & 0.66 \\
CosyVoice 3 & 0.72 \\
\midrule
Chroma 1.0 & \textbf{0.81}\footnotemark \\
\bottomrule
\end{tabular}
\caption{Performance comparison of speech models on zero-shot voice cloning. Higher SIM indicates better speaker similarity.}
\label{tab:wer_sim}
\end{table}
\footnotetext{Chroma operates at 24kHz sample rate, which better preserves speaker characteristics compared to 16kHz used by other models.}

\subsection{Objective Voice Cloning Evaluation}

We evaluate Chroma's voice cloning capability by measuring SIM, the primary focus of this work. Given that Chroma currently generates English-only audio, we assess its performance in a zero-shot setting using English samples from the CommonVoice dataset, following the protocol established by Seed-TTS~\cite{anastassiou2024seed}. Our model is evaluated at its native sampling rate of 24kHz. Table~\ref{tab:wer_sim} presents the comparative results against current state-of-the-art speech models and a human baseline.

As shown in Table~\ref{tab:wer_sim}, most contemporary TTS models achieve comparable speaker similarity scores, typically ranging 1--9\% below the human baseline. A notable exception is Seed-TTS, which slightly exceeds the human baseline. Chroma, on the other hand, outperforms all competing models, achieving a relative improvement of 10.96\% over the human baseline. This result suggests that Chroma effectively captures fine-grained paralinguistic features, enabling the generation of high-fidelity, personalized speech with exceptional speaker identity preservation.

\begin{table}[t]
\centering
\small
\begin{tabular}{lccc}
\toprule
\textbf{Metric} & \textbf{Chroma} & \textbf{ElevenLabs} & \textbf{Deuce}\\
\midrule
NCMOS & 24.4\% & \textbf{57.2\%} & 18.3\%\\
SCMOS & 40.6\% & 42.4\% & 17.0\%\\
\bottomrule
\end{tabular}
\caption{Comparative evaluation between Chroma and ElevenLabs.}
\label{tab:chroma_elevenlabs_comparison}
\end{table}

\begin{table}[t]
\centering
\small
\begin{tabular}{lcc}
\toprule
\textbf{Preference} & \textbf{Reference} & \textbf{ElevenLabs}\\
\midrule
Total & 4 (8.0\%) & \textbf{46 (92.0\%)} \\
\bottomrule
\end{tabular}
\caption{Comparison between ElevenLabs and reference (ground truth) audio.}
\label{tab:elevenlabs_vs_reference}
\end{table}

\begin{table*}[t]
\centering
\small
\begin{tabular}{lccc}
\toprule
\textbf{Component} &
\textbf{TTFT (ms)} &
\textbf{Avg Latency per Frame (ms)} &
\textbf{Total Duration (s)} \\
\midrule
Reasoner & 119.12 & 26.03 & 3.74 \\
Backbone & 8.48 & 8.75 & 4.27 \\
Decoder & 19.27 & 17.56 & 8.57 \\
Codec Decoder & -- & 3.08 & 2.99 \\
\midrule
Overall Generation Latency & \textbf{146.87} & 52.34 & \textbf{16.58} \\
Generated Audio Length & -- & -- & 38.80 \\
\midrule
\textbf{Generation RTF} & & \textbf{0.43} & \\
\bottomrule
\end{tabular}
\caption{Latency breakdown for speech generation. The system achieves TTFT of \textbf{146.87ms} and RTF of \textbf{0.43}.}
\label{tab:latency_breakdown}
\end{table*}

\subsection{Subject Voice Cloning Evaluation}
We conducted comparative experiments between Chroma and ElevenLabs, a state-of-the-art commercial voice cloning system. The experiments measured both NCMOS and SCMOS. We used their \texttt{eleven\_multilingual\_v2} API\footnote{\url{https://elevenlabs.io/docs/overview/models}} to generate outputs. For each dimension, we selected 15 samples and generated corresponding outputs using both models, resulting in 30 comparative samples evaluated across 12 independent sessions. Responses of ``About the same'' or ``Hard to tell'' were grouped into a ``Deuce'' category, representing cases where no clear preference could be established.

Table~\ref{tab:chroma_elevenlabs_comparison} presents the comparative results. For NCMOS, ElevenLabs demonstrated superior performance, receiving 57.2\% of preferences compared to Chroma's 24.4\% (18.3\% Deuce), indicating significantly more natural-sounding speech. However, for SCMOS, where participants first listened to reference audio before comparing synthesized outputs, the results were remarkably close: ElevenLabs received 42.4\% compared to Chroma's 40.6\% (17.0\% Deuce). This near-tie, with only 1.8 percentage points difference, suggests comparable capability in capturing speaker-specific characteristics.

This phenomenon reflects fundamental differences in system design. ElevenLabs employs a two-stage approach: first creating a voice profile from reference audio, then using it for TTS generation. In contrast, Chroma uses an end-to-end approach that directly processes reference audio in a single pass. While ElevenLabs optimizes for naturalness and clarity, its two-stage pipeline may lose fine-grained speaker characteristics during voice profile extraction. Chroma's end-to-end architecture preserves these nuanced features by maintaining direct access to reference audio throughout generation.

To investigate this trade-off, we conducted an additional experiment comparing ElevenLabs outputs directly with reference audio. Table~\ref{tab:elevenlabs_vs_reference} shows results from 5 sessions with 10 samples each, where evaluators chose which audio sounded more natural and human-like. Remarkably, evaluators overwhelmingly preferred ElevenLabs-generated audio (92.0\%) over actual human recordings (8.0\%). This reveals a critical insight: \textit{subjective listener preference does not necessarily align with speaker similarity}.

This finding reframes our SCMOS interpretation. While ElevenLabs achieved marginally higher SCMOS preference (42.4\% vs 40.6\%), this advantage likely reflects listeners' inherent bias toward naturalness rather than superior speaker fidelity. The overwhelming preference (92\%) for synthesized audio over ground truth recordings demonstrates that naturalness dominates speaker similarity in subjective evaluations. Consequently, Chroma's competitive SCMOS performance suggests stronger speaker fidelity than the raw percentages indicate, as it maintains competitiveness despite prioritizing faithful reproduction of speaker characteristics---including natural imperfections, speaking rate variations, and subtle prosodic features---over the perceptual naturalness that ElevenLabs optimizes for.

\begin{table*}[t]
\centering
\small
\setlength{\tabcolsep}{2pt}
\begin{tabular}{lccccccccccccc}
\toprule
\multirow{2}{*}{\textbf{Models}} & 
\multirow{2}{*}{\textbf{LLM Scale}} &
\multicolumn{3}{c}{\textbf{Understanding}} & 
\multicolumn{3}{c}{\textbf{Reasoning}} & 
\multicolumn{4}{c}{\textbf{Oral Conv.}} & 
\multirow{2}{*}{\textbf{Overall}} \\
\cmidrule(lr){3-5} \cmidrule(lr){6-8} \cmidrule(lr){9-12}
& & Rep.$\uparrow$ & Sum.$\uparrow$ & Gaokao$\uparrow$ & Storal$\uparrow$ & Truth.$\uparrow$ & Gsm8k$\uparrow$ & MLC$\uparrow$ & Alpaca$\uparrow$ & Common$\uparrow$ & Wild.$\uparrow$ & \\
\midrule
GLM-4-Voice & 9B & 90.95 & 91.07 & 64.47 & 73.80 & 59.28 & 30.93 & 57.82 & 80.77 & 63.07 & 78.76 & 69.09 \\
LLaMA-Omni & 8B & 45.62 & 80.68 & 16.06 & 50.65 & 45.13 & 3.89 & 44.44 & 64.36 & 58.40 & 72.19 & 48.14 \\
Freeze-Omni & 7B & 70.89 & 78.87 & 26.29 & 57.74 & 46.95 & 2.81 & 42.56 & 52.23 & 48.70 & 55.80 & 48.28 \\
\midrule
Mini-Omni & 0.5B & 5.07 & 32.20 & 0 & 23.25 & 25.06 & 0 & 2.82 & 30.99 & 29.80 & 31.42 & 18.06 \\
Mini-Omni2 & 0.5B & 8.10 & 40.06 & 0.66 & 28.49 & 26.92 & 0 & 6.97 & 34.81 & 30.70 & 36.43 & 21.31 \\
SLAM-Omni & 0.5B & 12.26 & 66.21 & 1.32 & 36.95 & 34.65 & 0 & 21.85 & 48.98 & 41.03 & 52.61 & 31.59 \\
\midrule
\textbf{Chroma (ours)} & \textbf{4B} & \textbf{69.05} & \textbf{74.12} & \textbf{38.61} & \textbf{71.14} & \textbf{51.69} & \textbf{22.74} & \textbf{60.26} & \textbf{60.47} & \textbf{62.07} & \textbf{64.24} & \textbf{57.44} \\
\bottomrule
\end{tabular}
\caption{Task accomplishment scores for end-to-end spoken dialogue models across understanding, reasoning, and oral conversation capabilities. Bold values indicate Chroma's performance, which remains competitive across all dimensions despite using only 4B parameters.}
\label{tab:e2e_task_scores}
\end{table*}

\subsection{Practical Generation Latency}

The current Chroma architecture does not support batch processing, therefore, we measured its latency under concurrency 1. Table~\ref{tab:latency_breakdown} presents the latency breakdown for each component during response generation. The total generation latency is the sum of all individual component latencies. We demonstrate the measured latency on a real example generating a 38.80-second audio response.

\paragraph{Prefilling Strategy.}
Before generating speech, we prefill the prompt text and prompt audio to reduce TTFT. Specifically, we encode and concatenate the prompt text and prompt audio to obtain prompt embeddings, which are then fed into the Backbone to perform prefill computation and generate the corresponding KV cache~\cite{pope2023efficiently}. By constructing the KV cache before the generation phase, the model avoids reprocessing prompt content during inference, enabling immediate autoregressive generation upon receiving user input. This strategy effectively reduces response latency and substantially improves real-time interaction performance.

\paragraph{Component-wise Latency Analysis.}
The Reasoner TTFT (119.12ms) represents the time from processing input data to generating the first audio token. The Backbone TTFT (8.48ms) measures the time to produce the first audio hidden states after receiving the audio token from the Reasoner. The Decoder then generates the remaining 7 codebooks ($c^{1:7}$) to capture fine-grained acoustic features, taking an average of 17.56ms per frame. Finally, the Codec Decoder reconstructs the waveform from the complete codebook sequence. Since we concatenate every 4 frames before passing them to the Codec Decoder for efficient batch processing, TTFT is not applicable for this component.

The overall system achieves a TTFT of 146.87ms, demonstrating sub-second responsiveness suitable for real-time interaction. The average latency per frame is 52.34ms, and with an RTF of 0.43, the system generates speech significantly faster than real-time playback, enabling smooth streaming generation.

\paragraph{Real-Time Factor.}
The generation RTF is calculated by dividing the total generation latency by the length of the generated audio:
\begin{equation}
\text{RTF} = \frac{T_{\text{generation}}}{T_{\text{audio}}} = \frac{16.58\text{s}}{38.80\text{s}} \approx 0.43.
\end{equation}
An RTF of 0.43 indicates that Chroma generates speech 2.3× faster than real-time playback, demonstrating strong performance for streaming applications. This efficiency allows the system to maintain low latency even during extended multi-turn conversations.

\subsection{Reasoning and Dialogue Capabilities}

While Chroma's primary focus is high-fidelity voice cloning, we evaluate its general dialogue capabilities to demonstrate that personalized voice generation does not compromise cognitive and conversational abilities. Table~\ref{tab:e2e_task_scores} compares Chroma against existing end-to-end spoken dialogue models across three dimensions using the basic track of \textbf{URO-Bench}~\cite{yan2025uro}.

Despite being optimized for voice cloning, Chroma demonstrates strong performance across all capabilities. In \textbf{reasoning tasks}, Chroma consistently achieves second-best performance: 71.14\% on Storal (vs. 73.80\% for GLM-4-Voice), 51.69\% on TruthfulQA (vs. 59.28\%), and 22.74\% on GSM8K (vs. 30.93\%). Notably, GLM-4-Voice uses 9B parameters, more than twice Chroma's 4B scale. In \textbf{oral conversation}, Chroma achieves the highest scores on MLC (60.26\%) and CommonVoice (62.07\%), demonstrating natural dialogue flow. For \textbf{understanding tasks}, Chroma achieves competitive scores of 69.05\% on repetition and 74.12\% on summarization, ranking second among all models.

Critically, Chroma is the \textit{only model in this comparison with personalized voice cloning capability}. All other models focus exclusively on dialogue and reasoning, without the ability to clone specific speaker characteristics. This makes Chroma's competitive performance particularly noteworthy: it maintains strong cognitive and conversational abilities while simultaneously supporting high-fidelity voice personalization, a capability absent in all compared systems. Furthermore, with only 4B parameters, Chroma achieves efficiency advantages over larger models (7B-9B) and delivers superior performance compared to smaller models (0.5B) across all evaluated tasks.

\section{Conclusion}

We present Chroma 1.0, the first open-source, real-time end-to-end spoken dialogue model that achieves both low-latency interaction and high-fidelity personalized voice cloning. Trained entirely on synthetic speech-to-speech data, Chroma achieves significant improvements in speaker similarity from only a few seconds of reference audio while maintaining real-time performance. Comparative evaluations demonstrate competitive performance against state-of-the-art commercial systems in voice cloning, alongside strong reasoning and dialogue capabilities. The system's efficient architecture enables smooth streaming generation suitable for natural multi-turn conversations. By combining low-latency streaming with high-fidelity voice cloning, Chroma opens new possibilities for accessible, personalized voice AI applications across diverse domains. We release our code and models to facilitate further research in personalized spoken dialogue systems.

\section{Acknowledgements}
We thank Yusen Lin, Kaiwen Zhou, and Liujie Zhen for their contributions to the early stages of this work. We are grateful to the volunteers who participated in our human evaluation experiments.

\bibliography{main}

@article{hsu2021hubert,
  title={Hubert: Self-supervised speech representation learning by masked prediction of hidden units},
  author={Hsu, Wei-Ning and Bolte, Benjamin and Tsai, Yao-Hung Hubert and Lakhotia, Kushal and Salakhutdinov, Ruslan and Mohamed, Abdelrahman},
  journal={IEEE/ACM transactions on audio, speech, and language processing},
  volume={29},
  pages={3451--3460},
  year={2021},
  publisher={IEEE}
}

@article{défossez2022highfidelityneuralaudio,
  title={High fidelity neural audio compression},
  author={D{\'e}fossez, Alexandre and Copet, Jade and Synnaeve, Gabriel and Adi, Yossi},
  journal={arXiv preprint arXiv:2210.13438},
  year={2022}
}

@article{ji2025wavtokenizerefficientacousticdiscrete,
  title={Wavtokenizer: an efficient acoustic discrete codec tokenizer for audio language modeling},
  author={Ji, Shengpeng and Jiang, Ziyue and Wang, Wen and Chen, Yifu and Fang, Minghui and Zuo, Jialong and Yang, Qian and Cheng, Xize and Wang, Zehan and Li, Ruiqi and others},
  journal={arXiv preprint arXiv:2408.16532},
  year={2024}
}

@article{défossez2024moshispeechtextfoundationmodel,
  title={Moshi: a speech-text foundation model for real-time dialogue},
  author={D{\'e}fossez, Alexandre and Mazar{\'e}, Laurent and Orsini, Manu and Royer, Am{\'e}lie and P{\'e}rez, Patrick and J{\'e}gou, Herv{\'e} and Grave, Edouard and Zeghidour, Neil},
  journal={arXiv preprint arXiv:2410.00037},
  year={2024}
}

@article{zeng2024glm4voiceintelligenthumanlikeendtoend,
  title={Glm-4-voice: Towards intelligent and human-like end-to-end spoken chatbot},
  author={Zeng, Aohan and Du, Zhengxiao and Liu, Mingdao and Wang, Kedong and Jiang, Shengmin and Zhao, Lei and Dong, Yuxiao and Tang, Jie},
  journal={arXiv preprint arXiv:2412.02612},
  year={2024}
}

@article{xu2025qwen25omnitechnicalreport,
  title={Qwen2. 5-omni technical report},
  author={Xu, Jin and Guo, Zhifang and He, Jinzheng and Hu, Hangrui and He, Ting and Bai, Shuai and Chen, Keqin and Wang, Jialin and Fan, Yang and Dang, Kai and others},
  journal={arXiv preprint arXiv:2503.20215},
  year={2025}
}

@misc{xu2025qwen3omnitechnicalreport,
      title={Qwen3-Omni Technical Report}, 
      author={Jin Xu and Zhifang Guo and Hangrui Hu and Yunfei Chu and Xiong Wang and Jinzheng He and Yuxuan Wang and Xian Shi and Ting He and Xinfa Zhu and Yuanjun Lv and Yongqi Wang and Dake Guo and He Wang and Linhan Ma and Pei Zhang and Xinyu Zhang and Hongkun Hao and Zishan Guo and Baosong Yang and Bin Zhang and Ziyang Ma and Xipin Wei and Shuai Bai and Keqin Chen and Xuejing Liu and Peng Wang and Mingkun Yang and Dayiheng Liu and Xingzhang Ren and Bo Zheng and Rui Men and Fan Zhou and Bowen Yu and Jianxin Yang and Le Yu and Jingren Zhou and Junyang Lin},
      year={2025},
      eprint={2509.17765},
      archivePrefix={arXiv},
      primaryClass={cs.CL},
      url={https://arxiv.org/abs/2509.17765}, 
}

@article{nguyen2024spiritlminterleavedspoken,
  title={Spirit-lm: Interleaved spoken and written language model},
  author={Nguyen, Tu Anh and Muller, Benjamin and Yu, Bokai and Costa-Jussa, Marta R and Elbayad, Maha and Popuri, Sravya and Ropers, Christophe and Duquenne, Paul-Ambroise and Algayres, Robin and Mavlyutov, Ruslan and others},
  journal={Transactions of the Association for Computational Linguistics},
  volume={13},
  pages={30--52},
  year={2025},
  publisher={MIT Press 255 Main Street, 9th Floor, Cambridge, Massachusetts 02142, USA~…}
}

@article{huang2025stepaudiounifiedunderstandinggeneration,
  title={Step-audio: Unified understanding and generation in intelligent speech interaction},
  author={Huang, Ailin and Wu, Boyong and Wang, Bruce and Yan, Chao and Hu, Chen and Feng, Chengli and Tian, Fei and Shen, Feiyu and Li, Jingbei and Chen, Mingrui and others},
  journal={arXiv preprint arXiv:2502.11946},
  year={2025}
}

@article{wu2025stepaudio2technicalreport,
  title={Step-audio 2 technical report},
  author={Wu, Boyong and Yan, Chao and Hu, Chen and Yi, Cheng and Feng, Chengli and Tian, Fei and Shen, Feiyu and Yu, Gang and Zhang, Haoyang and Li, Jingbei and others},
  journal={arXiv preprint arXiv:2507.16632},
  year={2025}
}

@article{huang2025stepaudioaqaafullyendtoendexpressive,
  title={Step-Audio-AQAA: a Fully End-to-End Expressive Large Audio Language Model},
  author={Huang, Ailin and Li, Bingxin and Wang, Bruce and Wu, Boyong and Yan, Chao and Feng, Chengli and Wang, Heng and Zhou, Hongyu and Wang, Hongyuan and Li, Jingbei and others},
  journal={arXiv preprint arXiv:2506.08967},
  year={2025}
}

@article{chu2024qwen2audiotechnicalreport,
  title={Qwen2-audio technical report},
  author={Chu, Yunfei and Xu, Jin and Yang, Qian and Wei, Haojie and Wei, Xipin and Guo, Zhifang and Leng, Yichong and Lv, Yuanjun and He, Jinzheng and Lin, Junyang and others},
  journal={arXiv preprint arXiv:2407.10759},
  year={2024}
}

@article{kimiteam2025kimiaudiotechnicalreport,
  title={Kimi-audio technical report},
  author={Ding, Ding and Ju, Zeqian and Leng, Yichong and Liu, Songxiang and Liu, Tong and Shang, Zeyu and Shen, Kai and Song, Wei and Tan, Xu and Tang, Heyi and others},
  journal={arXiv preprint arXiv:2504.18425},
  year={2025}
}

@article{wang2023neuralcodeclanguagemodels,
  title={Neural codec language models are zero-shot text to speech synthesizers},
  author={Wang, Chengyi and Chen, Sanyuan and Wu, Yu and Zhang, Ziqiang and Zhou, Long and Liu, Shujie and Chen, Zhuo and Liu, Yanqing and Wang, Huaming and Li, Jinyu and others},
  journal={arXiv preprint arXiv:2301.02111},
  year={2023}
}

@article{du2024cosyvoicescalablemultilingualzeroshot,
  title={Cosyvoice: A scalable multilingual zero-shot text-to-speech synthesizer based on supervised semantic tokens},
  author={Du, Zhihao and Chen, Qian and Zhang, Shiliang and Hu, Kai and Lu, Heng and Yang, Yexin and Hu, Hangrui and Zheng, Siqi and Gu, Yue and Ma, Ziyang and others},
  journal={arXiv preprint arXiv:2407.05407},
  year={2024}
}

@article{du2024cosyvoice2scalablestreaming,
  title={Cosyvoice 2: Scalable streaming speech synthesis with large language models},
  author={Du, Zhihao and Wang, Yuxuan and Chen, Qian and Shi, Xian and Lv, Xiang and Zhao, Tianyu and Gao, Zhifu and Yang, Yexin and Gao, Changfeng and Wang, Hui and others},
  journal={arXiv preprint arXiv:2412.10117},
  year={2024}
}

@article{du2025cosyvoice3inthewildspeech,
  title={Cosyvoice 3: Towards in-the-wild speech generation via scaling-up and post-training},
  author={Du, Zhihao and Gao, Changfeng and Wang, Yuxuan and Yu, Fan and Zhao, Tianyu and Wang, Hao and Lv, Xiang and Wang, Hui and Ni, Chongjia and Shi, Xian and others},
  journal={arXiv preprint arXiv:2505.17589},
  year={2025}
}

@article{xie2024miniomnilanguagemodelshear,
  title={Mini-omni: Language models can hear, talk while thinking in streaming},
  author={Xie, Zhifei and Wu, Changqiao},
  journal={arXiv preprint arXiv:2408.16725},
  year={2024}
}

@article{xie2024miniomni2opensourcegpt4ovision,
  title={Mini-omni2: Towards open-source gpt-4o with vision, speech and duplex capabilities},
  author={Xie, Zhifei and Wu, Changqiao},
  journal={arXiv preprint arXiv:2410.11190},
  year={2024}
}

@article{fang2025llamaomniseamlessspeechinteraction,
  title={Llama-omni: Seamless speech interaction with large language models},
  author={Fang, Qingkai and Guo, Shoutao and Zhou, Yan and Ma, Zhengrui and Zhang, Shaolei and Feng, Yang},
  journal={arXiv preprint arXiv:2409.06666},
  year={2024}
}

@misc{communication2023seamlessm4tmassivelymultilingual,
  title={SeamlessM4t—massively multilingual and multimodal machine translation},
  author={Duquenne, Paul-Ambroise and Elsahar, H and Gong, H and Heffernan, K and Hoffman, J and Klaiber, C and others},
  year={2023},
  publisher={Menlo Park, California, United States: Meta}
}

@article{openai2024gpt4ocard,
  title={Gpt-4o system card},
  author={Hurst, Aaron and Lerer, Adam and Goucher, Adam P and Perelman, Adam and Ramesh, Aditya and Clark, Aidan and Ostrow, AJ and Welihinda, Akila and Hayes, Alan and Radford, Alec and others},
  journal={arXiv preprint arXiv:2410.21276},
  year={2024}
}

@article{zhang2023speakforeignlanguagesvoice,
  title={Speak foreign languages with your own voice: Cross-lingual neural codec language modeling},
  author={Zhang, Ziqiang and Zhou, Long and Wang, Chengyi and Chen, Sanyuan and Wu, Yu and Liu, Shujie and Chen, Zhuo and Liu, Yanqing and Wang, Huaming and Li, Jinyu and others},
  journal={arXiv preprint arXiv:2303.03926},
  year={2023}
}

@article{qin2024openvoiceversatileinstantvoice,
  title={Openvoice: Versatile instant voice cloning},
  author={Qin, Zengyi and Zhao, Wenliang and Yu, Xumin and Sun, Xin},
  journal={arXiv preprint arXiv:2312.01479},
  year={2023}
}

@article{li2023styletts2humanleveltexttospeech,
  title={Styletts 2: Towards human-level text-to-speech through style diffusion and adversarial training with large speech language models},
  author={Li, Yinghao Aaron and Han, Cong and Raghavan, Vinay and Mischler, Gavin and Mesgarani, Nima},
  journal={Advances in Neural Information Processing Systems},
  volume={36},
  pages={19594--19621},
  year={2023}
}

@article{ju2024naturalspeech3zeroshotspeech,
  title={Naturalspeech 3: Zero-shot speech synthesis with factorized codec and diffusion models},
  author={Ju, Zeqian and Wang, Yuancheng and Shen, Kai and Tan, Xu and Xin, Detai and Yang, Dongchao and Liu, Yanqing and Leng, Yichong and Song, Kaitao and Tang, Siliang and others},
  journal={arXiv preprint arXiv:2403.03100},
  year={2024}
}

@article{le2023voiceboxtextguidedmultilingualuniversal,
  title={Voicebox: Text-guided multilingual universal speech generation at scale},
  author={Le, Matthew and Vyas, Apoorv and Shi, Bowen and Karrer, Brian and Sari, Leda and Moritz, Rashel and Williamson, Mary and Manohar, Vimal and Adi, Yossi and Mahadeokar, Jay and others},
  journal={Advances in neural information processing systems},
  volume={36},
  pages={14005--14034},
  year={2023}
}

@inproceedings{lewis2019bartdenoisingsequencetosequencepretraining,
  title={BART: Denoising sequence-to-sequence pre-training for natural language generation, translation, and comprehension},
  author={Lewis, Mike and Liu, Yinhan and Goyal, Naman and Ghazvininejad, Marjan and Mohamed, Abdelrahman and Levy, Omer and Stoyanov, Veselin and Zettlemoyer, Luke},
  booktitle={Proceedings of the 58th annual meeting of the association for computational linguistics},
  pages={7871--7880},
  year={2020}
}

@article{raffel2023exploringlimitstransferlearning,
  title={Exploring the limits of transfer learning with a unified text-to-text transformer},
  author={Colin, Raffel},
  journal={J. Mach. Learn. Res.},
  volume={21},
  year={2020}
}

@article{yu2022scalingautoregressivemodelscontentrich,
  title={Scaling autoregressive models for content-rich text-to-image generation},
  author={Yu, Jiahui and Xu, Yuanzhong and Koh, Jing Yu and Luong, Thang and Baid, Gunjan and Wang, Zirui and Vasudevan, Vijay and Ku, Alexander and Yang, Yinfei and Ayan, Burcu Karagol and others},
  journal={arXiv preprint arXiv:2206.10789},
  volume={2},
  number={3},
  pages={5},
  year={2022}
}

@article{ouyang2022traininglanguagemodelsfollow,
  title={Training language models to follow instructions with human feedback},
  author={Ouyang, Long and Wu, Jeffrey and Jiang, Xu and Almeida, Diogo and Wainwright, Carroll and Mishkin, Pamela and Zhang, Chong and Agarwal, Sandhini and Slama, Katarina and Ray, Alex and others},
  journal={Advances in neural information processing systems},
  volume={35},
  pages={27730--27744},
  year={2022}
}

@article{rafailov2024directpreferenceoptimizationlanguage,
  title={Direct preference optimization: Your language model is secretly a reward model},
  author={Rafailov, Rafael and Sharma, Archit and Mitchell, Eric and Manning, Christopher D and Ermon, Stefano and Finn, Chelsea},
  journal={Advances in neural information processing systems},
  volume={36},
  pages={53728--53741},
  year={2023}
}

@inproceedings{huang2024audiogpt,
  title={Audiogpt: Understanding and generating speech, music, sound, and talking head},
  author={Huang, Rongjie and Li, Mingze and Yang, Dongchao and Shi, Jiatong and Chang, Xuankai and Ye, Zhenhui and Wu, Yuning and Hong, Zhiqing and Huang, Jiawei and Liu, Jinglin and others},
  booktitle={Proceedings of the AAAI Conference on Artificial Intelligence},
  volume={38},
  number={21},
  pages={23802--23804},
  year={2024}
}

@article{lin2024advancing,
  title={Advancing large language models to capture varied speaking styles and respond properly in spoken conversations},
  author={Lin, Guan-Ting and Chiang, Cheng-Han and Lee, Hung-yi},
  journal={arXiv preprint arXiv:2402.12786},
  year={2024}
}

@inproceedings{ao2022speecht5,
  title={Speecht5: Unified-modal encoder-decoder pre-training for spoken language processing},
  author={Ao, Junyi and Wang, Rui and Zhou, Long and Wang, Chengyi and Ren, Shuo and Wu, Yu and Liu, Shujie and Ko, Tom and Li, Qing and Zhang, Yu and others},
  booktitle={Proceedings of the 60th annual meeting of the association for computational linguistics (volume 1: Long papers)},
  pages={5723--5738},
  year={2022}
}

@article{touvron2023llama,
  title={Llama: Open and efficient foundation language models},
  author={Touvron, Hugo and Lavril, Thibaut and Izacard, Gautier and Martinet, Xavier and Lachaux, Marie-Anne and Lacroix, Timoth{\'e}e and Rozi{\`e}re, Baptiste and Goyal, Naman and Hambro, Eric and Azhar, Faisal and others},
  journal={arXiv preprint arXiv:2302.13971},
  year={2023}
}

@article{tan2024naturalspeech,
  title={Naturalspeech: End-to-end text-to-speech synthesis with human-level quality},
  author={Tan, Xu and Chen, Jiawei and Liu, Haohe and Cong, Jian and Zhang, Chen and Liu, Yanqing and Wang, Xi and Leng, Yichong and Yi, Yuanhao and He, Lei and others},
  journal={IEEE Transactions on Pattern Analysis and Machine Intelligence},
  volume={46},
  number={6},
  pages={4234--4245},
  year={2024},
  publisher={IEEE}
}

@article{chu2023qwen,
  title={Qwen-audio: Advancing universal audio understanding via unified large-scale audio-language models},
  author={Chu, Yunfei and Xu, Jin and Zhou, Xiaohuan and Yang, Qian and Zhang, Shiliang and Yan, Zhijie and Zhou, Chang and Zhou, Jingren},
  journal={arXiv preprint arXiv:2311.07919},
  year={2023}
}

@article{zhang2023speechgpt,
  title={Speechgpt: Empowering large language models with intrinsic cross-modal conversational abilities},
  author={Zhang, Dong and Li, Shimin and Zhang, Xin and Zhan, Jun and Wang, Pengyu and Zhou, Yaqian and Qiu, Xipeng},
  journal={arXiv preprint arXiv:2305.11000},
  year={2023}
}

@inproceedings{radford2023robust,
  title={Robust speech recognition via large-scale weak supervision},
  author={Radford, Alec and Kim, Jong Wook and Xu, Tao and Brockman, Greg and McLeavey, Christine and Sutskever, Ilya},
  booktitle={International conference on machine learning},
  pages={28492--28518},
  year={2023},
  organization={PMLR}
}

@article{chen2022wavlm,
  title={Wavlm: Large-scale self-supervised pre-training for full stack speech processing},
  author={Chen, Sanyuan and Wang, Chengyi and Chen, Zhengyang and Wu, Yu and Liu, Shujie and Chen, Zhuo and Li, Jinyu and Kanda, Naoyuki and Yoshioka, Takuya and Xiao, Xiong and others},
  journal={IEEE Journal of Selected Topics in Signal Processing},
  volume={16},
  number={6},
  pages={1505--1518},
  year={2022},
  publisher={IEEE}
}

@inproceedings{chen2022large,
  title={Large-scale self-supervised speech representation learning for automatic speaker verification},
  author={Chen, Zhengyang and Chen, Sanyuan and Wu, Yu and Qian, Yao and Wang, Chengyi and Liu, Shujie and Qian, Yanmin and Zeng, Michael},
  booktitle={ICASSP 2022-2022 IEEE International Conference on Acoustics, Speech and Signal Processing (ICASSP)},
  pages={6147--6151},
  year={2022},
  organization={IEEE}
}

@inproceedings{ardila2020common,
  title={Common voice: A massively-multilingual speech corpus},
  author={Ardila, Rosana and Branson, Megan and Davis, Kelly and Kohler, Michael and Meyer, Josh and Henretty, Michael and Morais, Reuben and Saunders, Lindsay and Tyers, Francis and Weber, Gregor},
  booktitle={Proceedings of the twelfth language resources and evaluation conference},
  pages={4218--4222},
  year={2020}
}

@article{loshchilov2017decoupled,
  title={Decoupled weight decay regularization},
  author={Loshchilov, Ilya and Hutter, Frank},
  journal={arXiv preprint arXiv:1711.05101},
  year={2017}
}

@article{anastassiou2024seed,
  title={Seed-tts: A family of high-quality versatile speech generation models},
  author={Anastassiou, Philip and Chen, Jiawei and Chen, Jitong and Chen, Yuanzhe and Chen, Zhuo and Chen, Ziyi and Cong, Jian and Deng, Lelai and Ding, Chuang and Gao, Lu and others},
  journal={arXiv preprint arXiv:2406.02430},
  year={2024}
}

@article{yan2025uro,
  title={Uro-bench: A comprehensive benchmark for end-to-end spoken dialogue models},
  author={Yan, Ruiqi and Li, Xiquan and Chen, Wenxi and Niu, Zhikang and Yang, Chen and Ma, Ziyang and Yu, Kai and Chen, Xie},
  journal={arXiv preprint arXiv:2502.17810},
  year={2025}
}

@article{pope2023efficiently,
  title={Efficiently scaling transformer inference},
  author={Pope, Reiner and Douglas, Sholto and Chowdhery, Aakanksha and Devlin, Jacob and Bradbury, James and Heek, Jonathan and Xiao, Kefan and Agrawal, Shivani and Dean, Jeff},
  journal={Proceedings of machine learning and systems},
  volume={5},
  pages={606--624},
  year={2023}
}

\newpage\appendix
\section*{Appendix}
\section{Limitations and Future Work}

The current system does not incorporate external tool use or task-specific post-training techniques such as reinforcement learning from human feedback (RLHF)~\cite{ouyang2022traininglanguagemodelsfollow} or direct preference optimization (DPO)~\cite{rafailov2024directpreferenceoptimizationlanguage}. Integrating such methods could further improve dialogue quality, instruction following, and user preference alignment, particularly for conversational naturalness and contextual appropriateness.

Recent work on multi-codebook token prediction (MTP), such as Qwen3-Omni~\cite{xu2025qwen3omnitechnicalreport}, reduces first-packet latency by predicting residual codebooks in parallel rather than sequentially. Investigating whether MTP can be integrated into Chroma's decoder stage without compromising voice cloning fidelity presents a promising avenue for further latency reduction.

Although Chroma's speech reasoner supports multilingual input (currently Chinese and English), the system generates speech output only in English. Extending codec training and decoder modules to support multilingual output generation would broaden the system's applicability across diverse linguistic contexts. Cross-lingual voice cloning, where input and output languages differ while preserving speaker identity, remains an important research direction.

While Chroma's Backbone employs a decoder-only architecture aligning with recent trends in speech language modeling, encoder-decoder architectures have demonstrated strong performance across multiple domains, including machine translation~\cite{raffel2023exploringlimitstransferlearning}, text generation~\cite{lewis2019bartdenoisingsequencetosequencepretraining}, text-to-image generation~\cite{yu2022scalingautoregressivemodelscontentrich}, and spoken language processing~\cite{ao2022speecht5}. These architectures may offer distinct advantages in controllability, cross-modal alignment, and explicit separation of understanding and generation processes. Exploring encoder-decoder designs for speech-to-speech dialogue could provide complementary benefits to our current approach, particularly for scenarios requiring fine-grained control over both semantic content and acoustic properties in voice cloning and streaming generation.
\section{Ethical Considerations}

While Chroma demonstrates substantial advancements in voice cloning technology, it also introduces important ethical considerations. The capability to generate high-fidelity, personalized speech from minimal reference audio raises several risks, including the potential for impersonation, fraud, and the creation of misleading or harmful audio content without an individual's consent.

To mitigate these risks, we recommend adopting strong technical and policy-based safeguards, such as:
\begin{itemize}
    \item Requiring explicit and verifiable consent for any form of voice cloning,
    \item Developing and deploying reliable synthetic speech detection mechanisms,
    \item Enforcing clear usage policies and access controls to prevent misuse,
    \item Investigating watermarking or traceability techniques for generated audio.
\end{itemize}

When developed and used responsibly, voice cloning systems like Chroma can provide meaningful benefits, such as supporting voice restoration for individuals with speech impairments, enabling inclusive accessibility applications, and facilitating creative or personalized communication tools. We encourage the research community to advance ethical guidelines, regulatory frameworks, and robust detection technologies in parallel with progress in speech synthesis systems to ensure their safe and responsible deployment.

\section{Training Objective}~\label{app:training_obj}

During training, the Reasoner is frozen and functions solely as a
feature provider. For each training pair consisting of an audio sample
$\mathbf{x}_{\text{audio}}$ and its corresponding transcription
$\mathbf{x}_{\text{text}}$, the Reasoner produces text embeddings
$\mathbf{E}_{\text{text}} \in \mathbb{R}^{T \times d}$ and multimodal
hidden states $\mathbf{H}_{\text{reasoner}} \in \mathbb{R}^{T \times d}$,
where $T$ is the length of the text sequence and $d$ is the hidden dimension. These representations remain fixed throughout optimization
and serve as semantic and prosodic conditioning for the downstream
acoustic models.

Conditioned on reference audio $\mathbf{x}_{\text{ref-audio}}$ and reference text $\mathbf{x}_{\text{ref-text}}$, the Backbone
autoregressively predicts a coarse discrete acoustic code sequence
\[
\mathbf{c}^{0} = \{ c_t^{0} \mid t = 1,\ldots,L \},
\qquad
c_t^{0} \in \{1,\ldots,V\},
\]
where $L$ denotes the number of audio frames and $V$ is the codebook size.

The Chroma Decoder further refines each frame by producing the remaining
$N-1$ residual quantization levels of an RVQ representation. For each
level $j \in \{1,\ldots,N-1\}$, the Decoder predicts
\[
\mathbf{c}^{j}
    = \{ c_t^{j} \mid t = 1,\ldots,L \},
\qquad
c_t^{j} \in \{1,\ldots,V\}.
\]
We define the complete $N$-level discrete acoustic representation as

\[
c_t^{0:N-1} = (c_t^{0}, c_t^{1}, \ldots, c_t^{N-1}).
\]

\paragraph{Backbone Loss.}

To maintain causal alignment between text and audio generation, the
Backbone at time $t$ is allowed to attend only to prefix information.
We define the causal conditioning set
\[
\mathbf{z}_{<t}
    =
    \big(
        c^{0}_{<t},\;
        \mathbf{E}_{\text{text},<t},\;
        \mathbf{H}_{\text{reasoner},<t}
    \big),
\]
where
$c^{0}_{<t}$
denotes the autoregressively generated acoustic prefix and
$\mathbf{E}_{\text{text},<t}$,
$\mathbf{H}_{\text{reasoner},<t}$
represent the Reasoner-provided prefix signals.

The Backbone predicts the coarse code sequence
$\mathbf{c}^{0} = \{c_t^{0}\}_{t=1}^{L}$,
and its loss is defined as
\begin{equation}
\mathcal{L}_{\text{backbone}}
    =
    -\frac{1}{L}
    \sum_{t=1}^{L}
    \log
    p\!\left(
        c_t^{0}
        \mid
        \mathbf{z}_{<t}
    \right).
\label{eq:backbone_loss}
\end{equation}

This objective trains the Backbone to capture inter-frame temporal
structure and produce a coarse acoustic representation aligned with the
text progression.

\paragraph{Decoder Loss.}

Given the coarse code $c_t^{0}$ and the Backbone hidden state
$\mathbf{h}_t^{B}$, the Chroma Decoder refines each frame $t$ via an
\emph{intra-frame} autoregressive process. Specifically, it predicts the
residual quantization levels
\[
c_t^{1:N-1}
    = (c_t^{1}, c_t^{2}, \ldots, c_t^{N-1}),
\qquad
c_t^{j} \in \{1,\ldots,V\}.
\]

The conditional distribution over refinement levels factorizes as
\begin{equation}
p_{\theta}\!\left(
    c_t^{1:N-1}
    \mid
    c_t^{0},\,\mathbf{h}_t^{B}
\right)
=
\prod_{j=1}^{N-1}
p_{\theta}\!\left(
    c_t^{j}
    \mid
    c_t^{0},\,
    \mathbf{h}_t^{B},\,
    c_t^{1:j-1}
\right),
\label{eq:decoder_factorization}
\end{equation}
where $c_t^{1:j-1} = (c_t^{1},\ldots,c_t^{j-1})$ denotes the prefix for
teacher forcing.

The training objective for the Chroma Decoder is the negative
log-likelihood across all frames and refinement levels:
\begin{equation}
\mathcal{L}_{\text{decoder}}
    =
    -\frac{1}{L}
    \sum_{t=1}^{L}
    \sum_{j=1}^{N-1}
    \log
    p_{\theta}\!\left(
        c_t^{j}
        \mid
        c_t^{0},\,
        \mathbf{h}_t^{B},\,
        c_t^{1:j-1}
    \right).
\label{eq:decoder_loss}
\end{equation}

This objective encourages the Decoder to progressively refine each
frame's acoustic representation while remaining consistent with the
Backbone coarse code $c_t^{0}$ and the contextual information encoded in
$\mathbf{h}_t^{B}$.

\subsection{Training Strategy}
We adopt a two-stage training strategy to stabilize optimization and progressively strengthen the refinement of acoustic representations.

In the first stage, we jointly train the Backbone and Decoder by setting the loss weight to $\lambda = 0.5$. This balanced weighting encourages the model to learn both the coarse acoustic code distribution and the residual quantization levels, helping the system establish consistent semantic-acoustic alignment in the early phase of training.

In the second stage, we freeze the Backbone parameters and set $\lambda = 1$, thereby shifting optimization entirely to the Decoder. This fine-tuning phase focuses on refining the higher-level quantization layers, enabling the model to capture fine-grained speech characteristics such as timbre nuances, prosodic variations, and articulatory details. As a result, the final model achieves improved voice cloning fidelity and enhanced overall speech naturalness.

\end{document}